# Bonding Nature, Structural Optimization, and Energetics studies of SiC Graphitic-Like layer Structures and Single/Double Walled Nanotubes

Ming Yu, C. S. Jayanthi, and S. Y. Wu
*University of Louisville, Department of Physics, Louisville, KY 40292*

Abstract

The structural optimization and energetics studies of SiC graphitic-like structures have been investigated theoretically in the context of formations of stable graphitic-like layer structures, single- and multi-walled nanotubes using the DFT-based Vienna *ab-inito* simulation package. The bonding nature of atoms in the optimized structures has been examined using a local analysis technique based on a self-consistent and environment-dependent semi-empirical Hamiltonian. Results of our studies reveal that stabilized SiC graphitic-like layer structures possess the $sp^2$ bonding nature, different from the $sp^3$ bonding nature in bulk SiC. Such flexibility in bonding configurations between Si and C atoms holds the possibility for a wide range of stable SiC-based structures, similar to those for carbon-based structures. In the case of SiC-based nanotubes, we have calculated quantities such as the strain energy, the degree of buckle in the cylindrical shell, and bond charges between Si and C atoms, to obtain an understanding of the optimized structures. The optimized interlayer spacing of SiC graphitic-like multilayer sheets has been found to depend on the ordering of atoms in different layers of the SiC graphitic-like structure (~3.7 Å for the Si-C sequence of bilayer arrangement versus ~4.8 Å for either the Si-Si or the C-C sequence of bilayer arrangement). These observations may be attributed to the Coulomb interactions due to the charge redistribution among Si and C atoms. On the other hand, the existence of two different ranges of interlayer separation in SiC double-walled nanotubes (~3.8 Å for zigzag and ~4.8 Å for armchair) is found to be related to whether the dominant interlayer neighbors are of the Si-C type or the Si-Si and C-C types. This conclusion appears also to be the underlying reason for the experimental observation of two ranges of interlayer separation in SiC multiwalled nanotubes.

PACS numbers: 61.46.-w, 61.46.Fg, 68.65.-k, 68.65.Ac, 81.07.-b, 81.07.De

**I. INTRODUCTION**

Bulk silicon carbide (SiC) has about 250 polytypes depending on the stacking arrangements along the hexagonal c-axis with sequences of SiC bilayers. The bonding nature between Si and C in these polytypes is of $sp^3$-type hybridization, a typical bonding nature in a tetrahedral network. Since those polytypes possess large energy band gaps, large thermal conductivities, and large breakdown electric fields, these unique physical and electronic properties make them suitable materials for the fabrication of electronic devices for high-temperature, high-power, high-voltage, and high-frequency applications[1] It is expected that such outstanding properties of the bulk SiC can be enhanced or altered in SiC-based quasi-one (two) dimensional structures. Recently, several experimental groups have successfully synthesized quasi-one dimensional SiC-based structures.[2-5] Those quasi-one dimensional SiC structures include SiC nanowires,[2-4] β-SiC (cubic zinc-blende structure) nanotubes (NT),[5] and SiC multiwalled nanotubes (MWNTs).[2] It was reported that the synthesized SiC nanowires are either along (111) orientation of β-SiC[2-3] or along (0001) orientation of α-SiC[4] (hexagonal wurtzite structure) with a spacing of 2.5 Å, a typical spacing between two bilayers in bulk SiC. The same spacing is also found in the synthesized β-SiC nanotubes.[5] But the interlayer separations of synthesized SiC MWNTs[2] are found to be about 3.8 Å and 4.2-4.5 Å, respectively. Such big spacing in SiC MWNTs implies that SiC MWNTs possess a graphitic-like structure, which is quite different from β- or α-SiC nanowires and bulk phases where the $sp^3$ hybridization between Si and C atoms dominates. Thus, the discovery of SiC MWNTs raises an interesting issue on the structure versus the bonding nature in SiC graphitic-like structures.

The fact that the interlayer spacing of SiC MWNT (in the range of 3.8-4.5 Å) is larger than that of β-SiC NWs suggests that the structure of such SiC MWNTs could be related to the graphitic-like structure, with the Si-C bond possibly exhibiting the $sp^2$-like feature. On the other hand, since the Si-C bonds in bulk SiC possess $sp^3$ bonding nature, it is also expected that the SiC graphitic-like structure may be stabilized to a buckling structure to reflect the $sp^3$ bonding feature. A recent Density Functional Theory (DFT) study on the strain energy and the electronic structure of SiC NT has indeed shown that the optimized SiC NT has a weakly buckled structure when the diameter is within 1.6 nm.[6] It is therefore important to ascertain various possible aspects of the bonding nature of the Si-C bond for a comprehensive understanding of the structural and other properties of SiC based systems. Thus the relevant questions to be raised must include: (1) what kinds of bonding nature exist in stabilizing SiC graphitic-like structures such as one-dimensional SiC nanotubes and two-dimensional SiC graphitic-like sheet? and (2) why does the interlayer spacing of the SiC MWNTs exhibit two ranges of values at ~3.8 Å and ~4.5 Å, while carbon MWNTs exhibit the interlayer spacing only around 3.4 Å?

In order to answer such questions, we carried out a comprehensive study on the bonding nature, the structural optimization, and the energetics of SiC graphitic-like structures using the DFT-based Vienna *ab-initio* simulation package (VASP).[7] We first study the stability of SiC graphitic-like sheet and found that the graphitic-like sheet is stabilized under a transition from the $sp^3$-like bonding in bulk SiC to the $sp^2$-like bonding between Si and C atoms. The results and detailed discussions are presented in section II. In section III, we rolled up the SiC graphitic-like sheet with an initial configuration of either a buckled or a flat sheet structure to form a SiC NT and performed structural optimization for such SiC NTs. The relaxed tube structure shows a slight buckling. Such buckling was found to strongly depend on the tube curvature, decreasing with increasing diameter, and disappearing in the limit of the graphitic-like sheet structure. In order to understand why the interlayer separation of SiC MWNTs exhibits two different ranges of spacing at ~3.8 Å and ~4.5 Å, respectively, we performed structural relaxations on (1) multilayered SiC graphitic-like structures with 2- , 3-, or 4-bilayers and (2) armchair and zigzag SiC double walled NTs (DWNTs) in section IV. The multilayered graphitic-like SiC structures were stabilized with the interlayer spacing in the neighborhood of ~3.7 Å or ~4.8 Å, depending on the order of SiC-bilayer sequence (Si-C or C-C sequence). This scenario is reinforced by the case studies of armchair and zigzag SiC DWNTs. The results are consistent with the experimental observation for SiC MWNTs,[2] indicating that the tube-like SiC NTs observed in experiments indeed have graphitic-like structures. Finally, we give our conclusion in section V.

## II. STRUCTURAL OPTIMIZATION AND THE CALCULATION OF ENERGETICS OF SIC GRAPHITIC-LIKE SHEET STRUCTURES

In our study on the structural optimization and energietics calculations for SiC graphitic-like structures, we performed a plane-wave basis DFT optimization using VASP.[7] The ultra-soft, gradient-corrected, Vanderbilt-type pseudo-potential (US-PP) was used for constructing the interaction between the valence electron and the atomic core, and the Perdew and Wang (PW1991) generalized gradient approximation (GGA) correction was used to treat the exchange-correlation. For the SiC graphitic-like sheet and multilayered structure optimization, we employed the two-dimensional periodic boundary condition in the layered plane with a vacuum region (15 Å) between sheets to ensure there was no interaction between SiC graphitic-like sheets. For the total energy calculation, the cut-off energy for the plane wave basis set was taken to be 287 eV, and several sets of k points were taken according to the Monkhorst-Pack scheme (*i.e.*, 26, 52, 186, and 512 k points) so as to ensure the convergence of the total energy. The energy convergence for the self-consistent calculation was set to $10^{-5}$ eV, and the structure was relaxed using a conjugate-gradient (CG) algorithm until the atomic force was less than $10^{-3}$ eV/Å.

We first examined the structural properties for various bulk phases of carbon (diamond and graphite with c/a ratio fixed at 2.726), Si (tetrahedral network), and SiC (zinc-blende and wurtzite), respectively. The optimized lattice constant are 3.560 Å for diamond, 2.462 Å for graphite, 5.450 Å for Si, 4.375 Å for zinc-blende SiC, and 3.090 Å for wurtzite SiC, respectively, which are within 0.3% of corresponding experimental results, *i.e.*, 3.567 Å for diamond,[8] 2.46 Å for graphite,[9] 5.431 Å for Si,[8] 4.360 Å for zinc-blende SiC,[10] and 3.076 Å for wurtzite SiC,[10] respectively. It was also found that the zinc-blende structure is slightly more stable than the wurtzite structure for bulk SiC (*i.e.*, the cohesive energies per atom are -7.401 eV for the zinc-blende and -7.396 eV for the wurtzite structure, respectively).

The tetrahedral symmetry in the stable bulk SiC indicates that Si and C atoms are favored to form *sp$^3$* hybridization. Could such bonding character also stabilize SiC graphitic-like structure? To answer this question, we intentionally started with an initial configuration of buckled SiC graphitic-like sheet with buckling of 0.63 Å (along z-axis) to allow the possible effect of the dangling bonds associated with Si atoms, as shown in the left panel of Fig. 1. To our surprise, this configuration eventually relaxed to a regular flat graphitic-like form with buckling less than $1.0 \times 10^{-3}$ Å (see the right panel of Fig. 1). The optimized SiC bond length in this flat graphitic-like structure is 1.78 Å (1.89 Å in our optimized bulk SiC). The corresponding cohesive energy per atom is 0.49 eV higher than that of bulk β-SiC. The amount of energy difference is similar to those found in group-III nitride graphitic-like structures,[11-13] suggesting the possibility of forming SiC graphitic-like sheet as a metastable or intermediate state under appropriate conditions. While the existence of the stable *flat* SiC graphitic-like structure already suggests that Si and C atoms may form *sp$^2$*-type bonding, an understanding of the underlying reason for the transition of the "unstable" buckled graphitic-like SiC structure to a "stable" flat graphitic-like sheet would shed more light on this scenario. For this purpose, we preformed an analysis of the local bonding charge and bonding energy[14] using the SCED-LCAO (self-consistent and environment dependent-linear combination of atomic orbitals) approach[15]. The parameter sets of the SCED-LCAO Hamiltonians for column IV elements, including Si and C, have been developed previously [15,16]. When applied to Si- and C-based structures, it has been shown that the corresponding SCED-LCAO Hamiltonians are reliable and transferable, thus possessing predictive power.[15,16] Because Si and C belong to the same column in the periodic table, their electronic structures are expected to be similar. Hence for the binary SiC SCED-LCAO Hamiltonian, we chose to construct its overlapping matrices $S$ and its scaling function $K$ by averaging the corresponding parameters defining the respective $S$ and $K$ for Si and C. Namely, the overlap matrix elements and the scaling function are defined as $S_{ij,\tau}^{a,b} = (A_\tau^{a,b} + B_\tau^{a,b} R_{ij})[1+e^{-\alpha_\tau^{a,b} d_\tau^{a,b}}]/[1+e^{-\alpha_\tau^{a,b}(d_\tau^{a,b}-R_{ij})}]$ and $K_{ij}^{a,b} = e^{\alpha_K^{a,b} R_{ij}}$, where $A_\tau^{a,b}$, $B_\tau^{a,b}$, $\alpha_\tau^{a,b}$, $d_\tau^{a,b}$, and $\alpha_K^{a,b}$ are parameters for binary system and are chosen as $A_\tau^{a,b} = (A_\tau^a + A_\tau^b)/2$, $B_\tau^{a,b} = (B_\tau^a + B_\tau^b)/2$, $\alpha_\tau^{a,b} = (\alpha_\tau^a + \alpha_\tau^b)/2$, $d_\tau^{a,b} = (d_\tau^a + d_\tau^b)/2$, and $\alpha_K^{a,b} = (\alpha_K^a + \alpha_K^b)/2$, respectively. Notice that $A_\tau^{element}$, $B_\tau^{element}$, $\alpha_\tau^{element}$, $d_\tau^{element}$, and $\alpha_K^{element}$ are corresponding parameters for single element, such as Si or C. Because of the definitions of the potential functions $V_N$ (electron-electron interaction) and $V_Z$, (electron-ion interaction) we constructed $V_N$ and $V_Z$ for SiC binary system by averaging the corresponding potentials of Si and C. Namely, $V_N^{a,b} = (V_N^a + V_N^b)/2$ and $V_Z^{a,b} = (V_Z^a + V_Z^b)/2$, where $V_N^{element}$ and $V_Z^{element}$ are corresponding potential functions for single element (for the definitions of the parameter functions $S$, $K$, $V_N$, and $V_Z$, see Ref. [15] for details). To validate the SCED-LCAO Hamiltonian constructed for SiC, we carried out a structural optimization for bulk SiC and confirmed the relative stability between the zinc-blende and the wurtzite SiC. The optimized lattice constants are 4.55 Å for the zinc-blende and 3.20 Å for the wurtzite structure, respectively. The results are consistent with our DFT calculations. We also obtained a flattened graphitic-like structure of SiC sheet with optimized lattice constant 3.14 Å from an initially buckled graphite-like sheet, similar to the result obtained by the DFT calculation mentioned above. We then use the SCED-LCAO Hamiltonian constructed for SiC to perform a local analysis. The results of the orbital bond charge between atomic sites $i$ and $j$ ($N_{ij}(\alpha)$, $\alpha=\sigma$, $\pi$), the ratio of the π orbital bond charge to the σ orbital bond charge ($N_{ij}(\pi)/N_{ij}(\sigma)$), the bonding energy ($-E_{ij}$) between atomic sites $i$ and $j$, and the charge on the atomic site $i$ ($N_i$) for the zinc-blende and buckled/flattened graphitic-like sheet structures are listed in Table 1, respectively (for the definitions of $N_{ij}(\alpha)$, $E_{ij}$, and $N_i$ refer to Ref. [14]). From Table 1, it is seen that the bond charge is distributed only on the $\sigma$ orbital in bulk β-SiC phase with no charge distribution on the π orbital, providing a clear picture of $sp^3$ hybridization in bulk *β*-SiC. But for graphitic-like sheet structures, there is charge distribution on the π orbital. Specifically the charge distribution on the π orbital undergoes a change from less than $0.02|e|$ per bond in the initial buckled and unrelaxed graphitic-like structure to about $0.10|e|$ per bond in the final stabilized and flattened graphitic-like sheet structure, leading to an enhancement of the ratio $N_{ij}(\pi)/N_{ij}(\sigma)$ from 0.054 to 0.30 (note that the ratio $N_{ij}(\pi)/N_{ij}(\sigma)$ for graphite is 0.33). The dangling bonds associated with the buckled and unrelaxed graphitic-like sheet structure are eliminated by increasing the charge distribution on the π orbital to form the $sp^2$ type of bonding. This scenario is reinforced by the examination of the bond energy ($-E_{ij}$). As shown in Table 1, the bonding energy of the relaxed flat graphitic-like structure is about 2.35 eV higher than that of the unrelaxed and buckled sheet structure which is still dominated by the $sp^3$-bonding. This then indicates that the energetics favors a transition from the dominant $sp^3$–bonding in the unrelaxed buckle sheet structure to the $sp^2$ bonding in the relaxed flat sheet structure. It is also an indication that the bonding between a silicon atom and a carbon atom, similar to the bonding between the carbon atoms, possesses a multi-facet nature. Namely, it could either be the $sp^3$-bonding as in the bulk SiC or the $sp^2$-bonding as in the graphitic-like sheet structure. Hence, even though Si atoms ordinarily favor the $sp^3$ hybridization, the presence of C atoms indeed has a stabilizing effect on the graphitic-like structure of SiC because a Si atom and a carbon atom could form a $sp^2$ type of bonding. This is in good agreement with the EELS spectra of a single SiC NT,[2] where a comparison to those of a single SiC NW indicates that the strong pre-edge adsorption peaks of SiC NT indicate a $\pi$ bonding between Si and C in SiC NT. These results clearly demonstrate that in a SiC graphitic-like structure, it is energetically favorable to form an alternative SiC bonding and the bonding is of the $sp^2$ type. Furthermore, the likelihood that a Si atom and a C atom may form either $sp^3$ bonding or $sp^2$ bonding holds promise for other potentially interesting and useful SiC-based structures.

From Table 1, it can also be seen that a considerable amount of electron charge is transferred from Si atomic site to C atomic site (i.e., 0.31|$e$| for bulk β-SiC and 0.32|$e$| for SiC graphitic-like structure, respectively). These values are close to the value of 0.45|$e$| for SiC NT obtained by using other *ab initio* calculations.[6] This is an indication that the bonding between a Si atom and a C atom in a SiC-based structure also carries a strong ionic bonding flavor.

**III. THE STRAIN ENERGY AND THE BUCKLING OF SIC NTs**

As described above, an initially buckled graphitic-like SiC sheet is found to stabilize to a flattened sheet. On the other hand, previous theoretical calculation showed that a SiC NT has a weakly buckled structure when its diameter is less then 1.6 nm.[6] Buckling is usually associated with the release of the strain energy. It is therefore interesting to examine how the buckling changes as a function of the diameter of the SiC NT and the role of the strain energy. For these purposes, we preformed structural optimization for armchair ($m$, 0) and zigzag ($m$, $m$) SiC NT structures of increasing diameters (*m* ranging from 5 to 20) using DFT-based VASP.[7] The calculation procedures for SiC NTs are the same as those for SiC graphitic-like sheet. The periodic boundary condition is employed for the quasi-one-dimensional SiC NT structure along the tube axis with a vacuum space (20 Å) between tubes to ensure no interaction between SiC NTs. The plane-wave cut-off is taken as 287 eV with 60 k points for SiC NTs, following the Monkhorst-Pack scheme.

Two kinds of initial configurations of SiC NTs were considered: one is rolled up from a graphitic-like sheet, and the other is rolled up from a buckled graphitic-like sheet with a buckling of 0.58 Å. The optimized SiC NTs with various diameters have the following characters which are almost independent of the two initial different configurations: (i) the charge transfer from a Si atom to a C atom is about 0.32|*e*|, and the average Si-C bond length of these tubes is about 1.78 Å, similar to the bond length of SiC graphitic-like sheet obtained in our calculations and in good agreement with other *ab inito* results for SiC NTs;[6,17-19] (ii) for the equilibrium configurations of SiC NTs, Si atoms move towards the tube axis while C atoms move in the opposite direction, resulting in a slightly buckled structure; (iii) the amount of the radial buckling ($\Delta r_{SWNT}$) only depends on the SiC NT diameter, but independent of the chirality of the tube, as shown in Fig. 2(a). Similar characteristics was also found for BN,[11] GaN,[12] and AlN[13] NTs, where the electronegative N atoms move outward from the tube axis while group-III atoms (B, Ga, Al) move inward to the tube axis. Since a C atom is electronegative in the SiC NT (gaining about 0.32|$e$| from Si atom), it acts like an N atom and the Si atom acts like a group-III atom, as expected.

The corresponding strain energy per atom of SiC NT ($E_{strain} = E_{tube} - E_{sheet}$, where $E_{tube}$ and $E_{sheet}$ represent the total energies per atom of the SiC NT and the SiC graphitic sheet, respectively) as a function of diameter is shown in Fig. 2(b). We also found that the strain energies are slightly lower compared to those of carbon NTs in good agreement with other DFT calculations.[19] This means that it will cost less energy to roll up a SiC NT from a SiC graphitic-like sheet than that required to form a carbon SWNT from a graphene. But because of the binding energy of a SiC sheet is 0.49 eV higher than that of the bulk SiC, the SiC NT is metastable and the SiC MWNTs could readily transform to cubic structures, such as $\beta$-SiC NW's as observed from the experiment.[2] From Figs 2(a) and 2(b), it can be seen that the buckling, as well as the strain energy, decreases rapidly with increasing tube diameter and disappears in the limit of the flattened graphitic-like sheet when the tube diameter goes to infinity. This unmistakable correlation between the buckling and the strain energy is an indication that the buckling is mainly due to the release of the strain energy induced by the curvature of the tube. In addition, it is found that the strain energy, determined from the difference between the binding energy of SiC NTs and that of the graphitic-like sheet, can be fitted as a function of the tube diameter by the expression $E_s=\alpha/D^2$, where $\alpha$=6.3 (eV Å$^2$), and $D$ is the diameter of the SiC NT.

**IV. EXAMINING INTERLAYER SPACING OF MULTILAYERED SIC GRAPHITIC-LIKE STRUCTURES AND SIC DUOBLE WALLED NANOTUBES**

In order to understand why the interlayer spacing of SiC MWNTs exhibits two ranges of values at ~3.8 Å and ~4.5 Å respectively, we first performed the structural optimization on several multilayered SiC graphitic-like structures (2-, 3-, and 4-bilayers) using the software package VASP[7]. In the optimization process, we chose strongly buckled initial configurations with the buckling up to 0.63 Å. Two sequences of bilayer arrangement were investigated. One is in the order of C-C arrangement with the C atoms in different bilayers ordered along the same line. The other is in the order of Si-C arrangement with the Si and C atoms in different bilayers ordered alternatively along the same line. Fig. 3 illustrates the relaxation of the 2-bilayers SiC graphitic-like structures with C-C or Si-C order, respectively. The structures in the left column of Fgi.3 are the initial buckled configurations with interlayer spacing values of 3.08 Å, 2.31 Å, and 1.54 Å, respectively. The structures in the right column of Fig.3 are the corresponding relaxed ones. We examined these structures with the initial interlayer spacing from 3.08 Å to 1.54 Å. After relaxation, all the buckled SiC multilayered samples are finally stabilized to a flattened multilayered structure with the interlayer spacing of ~ 3.66 Å for the Si-C order and 4.47 to 4.64 Å for the C-C order, respectively, as shown in Fig. 3 for 2-bilyaers cases. We have also examined multilayered structures cut from bulk β-SiC along the (111) direction and those cut from α-SiC along the (0001) direction, respectively. In these cases, we did not find any flattened and well separated graphitic multilayers after relaxation. The results clearly indicate that only the graphitic-like multilayer structure can stabilize with the interlayer spacing larger than 2.5 Å of the typical bilayer spacing in bulk SiC, and again confirm that the SiC MWNTs observed in experiments have the graphitic-like symmetry, not the cubic symmetry. The energies per atom of 2-bilayers (2BL) graphitic-like SiC structures relative to their corresponding values at infinite separation ( $\Delta E = E_{2BL} - E_{sheet}$ ), calculated for two different types of positional ordering (C-C and Si-C) between the layers, are shown as a function of the interlayer spacing in Fig. 4. It can be seen that the equilibrium interlayer distance is ~ 3.67 Å for the Si-C ordering and ~ 4.82 Å for the C-C ordering, respectively, close to the experimental observation of ~ 3.8 and ~ 4.5 Å of SiC MWNTs.[2] In contrast to the carbon based graphite with no charge redistribution on each site, each C atom gains 0.32|$e$| from each Si atom in the SiC graphitic-like sheet, resulting in long-range electrostatic Coulomb interactions not only between Si atoms and C atoms on the same bilayer but also between these atoms in different bilayers. Obviously, in addition to the weaker van der Waals interactions between C-C layers, the competition between the interlayer C-C/Si-Si repulsive interactions and the interlayer Si-C attractive interactions is an important factor in deciding the equilibrium interlayer spacing of the multilayered SiC graphitic-like structure. It is this competition that leads to a different range of equilibrium interlayer spacing for the case of the Si-C ordering where pairs of Si-C are the nearest interlayer neighbors as compared to that of the C-C/Si-Si ordering where pairs of C-C/Si-Si are the nearest interlayer neighbors. From Fig. 4, it can also be seen that the energy required to stabilize the multilayered SiC graphitic-like structure is about 1 meV for the C-C ordering and 6 meV for the Si-C ordering, respectively. This means that the dominant effects of the interlayer Si-C attractive interactions over the interlayer C-C/Si-Si repulsive interactions is stronger in the case of Si-C ordering, but weaker in the C-C/Si-Si ordering, leading to a more stable multilayer Si-C graphitic-like structure in the Si-C ordering than that in the C-C ordering.

Note however that there is no clear one-to-one Si-C or C-C ordering in multiwalled SiC nanotubes. Therefore, the competition between the interlayer C-C/Si-Si and the interlayer Si-C interactions becomes more complex. But it is clear that the interlayer spacing of multiwalled SiC nanotubes still depends on the combined effect of repulsive interactions between C (Si) and C (Si) atoms and attractive interactions between C and Si atoms in different layers. To shed light on how this combined effect affects the interlayer spacing of SiC MWNTs, we used the VASP[7] to examine the relaxations and the energetics of two sets of commensurate SiC DWNTs, namely the set of the zigzag nanotubes composed of (5,0)@($m$,0) (with $m$=12, 13, 14, 15, 17, 20) and the set of the armchair nanotubes composed of (5,5)@($m$,$m$) (with $m$=9, 10, 11, 12, 13, 14, 15). The relative energy/atom of the relaxed SiC DWNT with respect to corresponding SiC NTs ($\Delta E = E_{DWNT} - E_{SWNT-1} - E_{SWNT-2}$) is shown as a function of the optimized intershell spacing in Fig. 4 by squares (the armchair DWNTs) and full circles (the zigzag DWNTs), respectively. It can be seen that the two curves corresponding to the two sets of SiC DWNTs can be viewed as framed within the borders of the two curves of the 2-bilayer graphitic-like structures corresponding to the C-C/Si-Si order and Si-C order, respectively. Also, the curve for the zigzag DWNTs $(5,0)@(m,0)$ exhibits an energy minimum at ~ 3.8 Å while that for the armchair DWNTs $(5,5)@(m,m)$ exhibits an energy minimum at ~ 4.8 Å, consistent with the equilibrium interlayer separation exhibited by the 2-bilayer graphitic-like structure with the Si-C order and that exhibited by the structure with the C-C order, respectively. To shed light on this interesting correlation, we determined the ratio ($\alpha$) of the number of the nearest Si-C pairs to the total number of the nearest C-C pairs and Si-Si pairs for a given set of DWNTs within a range of separations of atomic pairs defined by its stabilized interlayer separation. Namely, $\alpha = n_{Si-C} / (n_{Si-Si} + n_{C-C})$, where $n_{Si-C}$ represents the numbers of Si-C pairs, $n_{C-C}$ the number of C-C pairs, and $n_{Si-Si}$ the number of Si-Si pairs, respectively. The ratio for each DWNT is indicated by the number in the parenthesis adjacent to the point representing the particular DWNT in Fig. 4. An examination of these ratios shows that the ratios for the zigzag DWNTs almost all exceed one with only one exception that is removed from the equilibrium interlayer separation (~ 3.8 Å) but still close to one while those for the armchair DWNTs and in the neighborhood of the equilibrium interlayer separation (~

4.8 Å) are substantially less than one. A ratio greater than one indicates that there are more Si-C pairs than the C-C and Si-Si pairs combined within the range under consideration while a ratio less than one indicates the opposite. Thus for the zigzag DWNTs in the neighborhood of the equilibrium interlayer separation where the ratios are greater than one, the dominance of Si-C pairs over the combined C-C and Si-Si pairs yields an equilibrium interlayer separation of ~ 3.8 Å while for the armchair DWNTs in the neighborhood of its equilibrium interlayer separation where the ratios are less than one, the dominance of the combined C-C and Si-Si pairs corresponds to an equilibrium interlayer separation of ~ 4.8 Å, consistent with the cases of the multi-layered graphitic-like structures discussed above. The result thus suggests that the nature of the interlayer neighbors is the key factor determining the equilibrium interlayer spacing. While it is computationally not feasible to determine the equilibrium interlayer spacing for general SiC MWNTs of any chirality because the structures of their shells are in general incommensurate, it is not unreasonable to suggest, based on the above consideration, that the experimentally observed two ranges of interlayer spacing of SiC MWNTs are the consequences of the competition between interlayer attractive and repulsive Coulomb interactions. Specifically, if the interlayer pairs are dominated by Si-C pairs, the equilibrium interlayer spacing will be in the range of ~ 3.8 Å while if the interlayer pairs are dominated by C-C and Si-Si pairs, the interlayer spacing will be in the range of ~ 4.8 Å.

We also examined the radial buckling feature of SiC DWNTs and the deviation of the radial buckling in SiC DWNTs ($\Delta r_{DWNT}$) with respect to that of isolated SiC NT ($\Delta r_{SWNT}$), as a function of the intershell spacing. As shown in Fig. 5, when the intershell spacing is less than 5 Å the buckling feature of the inner shell of SiC DWNTs is weakened while that of outer shell is enhanced, compared to the corresponding isolated SiC NT. Such behavior is almost independent of the chirality of SiC DWNTs. This reveals that when two shells are close, the Coulomb interaction between neighboring shells can affect the buckling behavior of the tubes. When the intershell spacing increases, the effect of the Coulomb interaction will be weakened as can be seen from the zero deviation of the radial buckling in SiC DWNT with large intershell spacing shown in Fig. 5.

It is expected that the outstanding properties of bulk SiC can be enhanced or altered in the qausi-one (two) dimensional SiC based materials because the quantum confinement effect may play a role in these cases. To shed light on it, we examined the electronic band structures of the armchair and the zigzag SiC single/double wall NTs, as well as the SiC graphitic-like sheet structure. The corresponding energy gaps are summarized in Fig. 6. We found that the zigzag SiC SWNTs have a direct gap while the armchair SiC SWNTs have an indirect gap, just as in the case of a SiC graphitic-like sheet. For a given diameter of the SiC SWNT, the energy gap of the armchair tubes is larger than that of the zigzag tubes with similar diameters. We also found that the energy gap of the SiC SWNT increases with increasing diameter of the tube and almost saturate to the value of the SiC graphitic-like sheet (2.54 eV) when the diameter is larger than 20 Å. For SiC DWNT with its diameter defined by the diameter of its outer shell, we found that the energy gap is basically related to its inner shell. Furthermore, we found that the energy gap of the armchair SiC DWNTs (5,5)@($m,m$) is smaller than that of corresponding-isolated inner tube (5,5), and the zigzag SiC DWNTs (5,0)@($m$,0) almost show metallic behavior, with energy gap somewhat smaller than the corresponding inner tube (5,0). Such reduction indicates that the Coulomb interaction between the neighboring shells of the SiC DWNTs may affect not only the buckling behavior but also the electronic structure in the double walled SiC NTs.

**V. CONCLUSION**

We have carried out the structural optimization and the calculation of the energetics of SiC graphitic-like structures using the DFT-based VASP[7]. The study yields to two major findings. (1) The bonding nature between a Si atom and a C atom could be either the *sp*[3] type as existing in bulk β-SiC and α-SiC or the *sp*[2] type as existing in SiC graphitic-like sheet structures and SiC NTs. The capability of forming *sp*[3] bond as well as *sp*[2] bond between a Si atom and a C atom provides the SiC-based systems with the flexibility, similar to the C-based systems but not available for the pure Si-based systems, of forming potentially interesting and useful structures. (2) Because of the charge redistribution between the Si atom and the C atom in SiC graphitic-like multilayer structures and the SiC DWNTs, the interlayer Coulomb interactions are mainly responsible for the equilibrium interlayer spacing in SiC MWNTs, leading to two ranges of interlayer spacing as observed in the experiments, depending on whether the dominant nearest interlayer neighbors are of the Si-C type or the C-C (Si-Si) type. Our study on the energetics of SiC SWNTs has indicated a clear correlation among the strain energy, the degree of buckling in the cylindrical shell, and the charge redistribution between C and Si. Our study on the electronic structures of SiC single/double walled NTs also indicates two issues. (i) SiC SWNTs have semiconductor nature with direct energy gap in zigzag SWNTs and indirect energy gap in armchair SWNTs, respectively. The energy gap increases with increasing diameter of the tube and will saturate to that of SiC graphitic-like sheet. (ii) The energy gap of double walled SiC NTs is mainly dependent on the electronic structure of the inner shell, with some smaller effect from the intershell Coulomb interaction.

Acknowledgement

This work was supported by the Kentucky Science and Engineering Foundation under Grant No. KSEF-753-RDE-007.

Table 1. Local analysis of the orbital bond charge ($N_{ij}(\alpha)$, $\alpha=\sigma, \pi$), the ratio of the $\pi$ orbital bond charge to the $\sigma$ orbital bond charge ($N_{ij}(\pi)/N_{ij}(\sigma)$), the bonding energy ($-E_{ij}$), and the charge located on atomic site $i$ ($N_i$) for the zinc-blende and initial buckled/relaxed flattened graphitic-like sheet structures, respectively.

| Structure | $N_{ij}(\sigma)$ | $N_{ij}(\pi)$ | $N_{ij}(\pi)/N_{ij}(\sigma)$ | $E_{ij}$ (eV) | $N_i$ (atom at site $i$) |
|---|---|---|---|---|---|
| Zinc-blende | 0.29\|e\| | 0.00\|e\| | 0.00 | -7.41 | 4.31 (C) 3.69 (Si) |
| Initial buckled graphitic-like sheet | 0.31\|e\| | 0.02\|e\| | 0.05 | -8.07 | 4.19 (C) 3.81 (Si) |
| Relaxed flattened graphitic-like sheet | 0.32\|e\| | 0.10\|e\| | 0.30 | -10.42 | 4.32 (C) 3.68 (Si) |

**Figure captions**

**Fig. 1** (Color online) The side-view of the initial buckled configuration of the SiC graphitic-like structure (yellow: carbon atoms, blue: Si atoms) is shown on the left panel, where $b_{Si-C}$=1.89 Å and $z_{buckled}$=0.63 Å (see (b)). The final relaxed structure is shown on the right panel, where (c) and (d), correspond to the side and top views, respectively. The relaxed structure is nearly a flat sheet with $b_{Si-C}$ = 1.78 Å and $z_{buckled}$ = 0.0007 Å.

**Fig. 2** (Color online) (a) The average radial buckling ($\Delta r_{SWNT}$) as a function of the tube diameter is shown for optimized structures of SiC SWNTs (both for the armchair (*m,m*) and the zigzag (*m*,0) tubes). The results correspond to two types of initial configurations: (i) buckled initial configuration ($z_{buckled}$=0.58, represented by circles) and (ii) unbuckled initial configuration ($z_{buckled}$=0, represented by squares). The inset shows the optimized structures corresponding to (5,5) and (10,0) SiC SWNTs with $b_{Si-C}$=1.78 Å (yellow: carbon atoms, blue: Si atoms). The dotted line is shown as guidance. (b) The strain energy per atom ($E_{strain} = E_{tube} - E_{sheet}$) as a function of the tube diameter. The circles and squares represent strain energies per atom of the zigzag and the armchair SiC SWNTs, respectively, while the stars represent the strain energy

per atom of carbon SWNTs. The strain energy per atom can be fitted to the equation $E_s = \alpha/D^2$ (dotted curve), where $\alpha=6.3$ (eV Å$^2$) and $D$ (Å), the tube diameter.

**Fig. 3** (Color online) Schematic illustration of relaxed 2-bilayers (2BL) SiC graphitic-like structures corresponding to two different orderings: (a) C-C ordering (top panel) and (b) Si-C ordering (bottom panel). The two different orderings are indicated by the vertical arrows in the right column. The relaxed structures (shown in the right column, where $b_{si-c}$ = 1.78 Å, $Z_{buckled}$= 0.007 Å) were obtained from three different buckled initial configurations (the structures shown in the left column, where $b_{si-c}$ = 1.89 Å, $Z_{buckled}$= 0.63 Å) with initial interlayer spacing values of 3.08 Å, 2.31 Å, and 1.54 Å, respectively. For the case of the C-C ordering, these initial configurations stabilized to flattened 2-bilayer SiC graphitic-like structures with interlayer spacing of ~ 4.47 Å. Similarly, for the case of the Si-C ordering, the relaxed structures exhibited interlayer spacing of ~ 3.66 Å.

**Fig. 4** (Color online) The energies per atom of 2-bilayers (2BL) graphitic-like SiC structures relative to their corresponding values at infinite separation ($\Delta E = E_{2BL} - E_{sheet}$) are shown as a function of the interlayer spacing. The upright triangles, representing the results corresponding to the C-C ordering between bilayers, exhibit a minimum at ~ 4.82 Å, while the inverted triangles, representing the results corresponding to the Si-C ordering between bilayers, exhibit a minimum at ~ 3.67 Å.

Also shown in the figure are the results corresponding to the relative energy/atom of the relaxed armchair and zigzag SiC DWNTs with respect to their component SWNTs ($\Delta E = E_{DWNT} - E_{SWNT-1} - E_{SWNT-2}$), as a function of the optimized intershell spacing. The filled squares represent the results for the armchair SiC DWNTs corresponding to (5,5)@($m,m$) ($m$=9,10,11,12,13,14,15), respectively, and the filled circles denote the results for the zigzag SiC DWNTs corresponding to (5,0)@($m$,0), ($m$=12,13,14,15,17,20), respectively. The curve for $\Delta E$ versus the intershell spacing exhibits a minimum at ~ 3.8 Å for the zigzag SiC DWNTs, and at ~ 4.8 Å for the armchair SiC DWNTs. The value of $\alpha$ (see the text for its definition) is indicated by the number adjacent to the point representing the DWNT. The inset shows the optimized structures corresponding to (5,0)@(12,0) and (5,5)@(10,10) SiC DWNTs (yellow: carbon atoms, blue: Si atoms), respectively.

**Fig. 5** (Color online) The open (filled) squares represent differences between the radial buckling values of the outer shells (inner shells) of the armchair SiC DWNTs ($\Delta r_{DWNT}$) and those of the corresponding isolated SiC SWNTs ($\Delta r_{SWNT}$), calculated for (5,5)@($m,m$) tubes with $m$ = 9,10,11,12,13,14,15, as a function of the intershell spacing. The open(filled) circles represent similar quantities for the zigzag SiC DWNTs ((5,0)@($m$,0), $m$ = 12,13,14,15,17,20).

**Fig. 6** (Color online) The energy gaps as a function of the tube diameter for the armchair (filled squares) and the zigazag (filled circles) SiC DWNTs which are compared with the corresponding results for SiC SWNTs (open squares for the armchair and open circles for the zigzag tubes, respectively). The energy gap of the SiC graphitic-like sheet is around ~ 2.54 eV (indicated by the star). The chiral indices of some SiC SWNTs/DWNTs are indicated at their corresponding energy gaps.

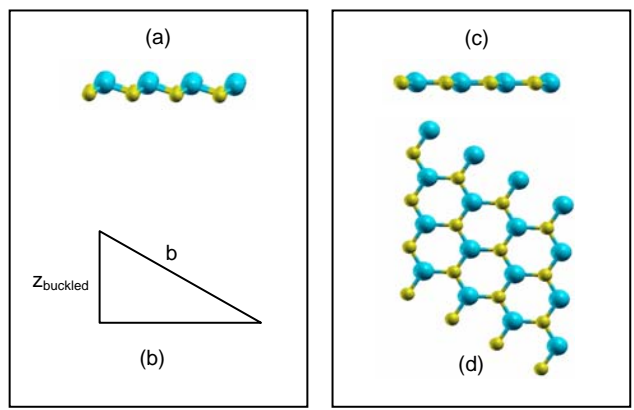

Figure 1

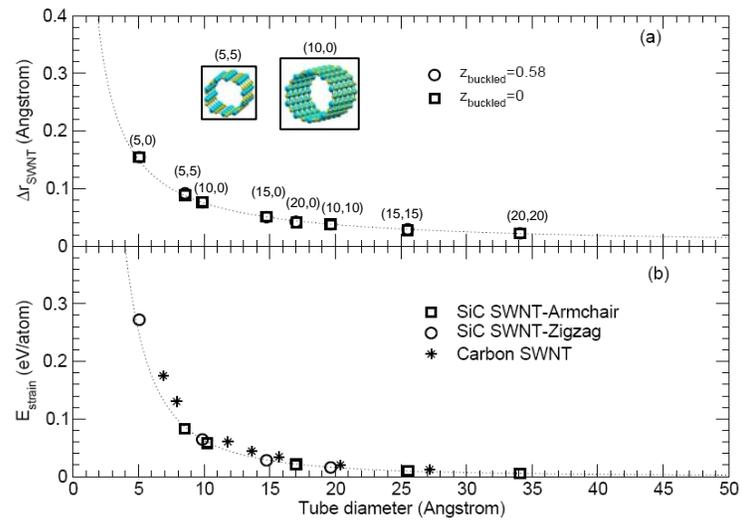

Figure 2

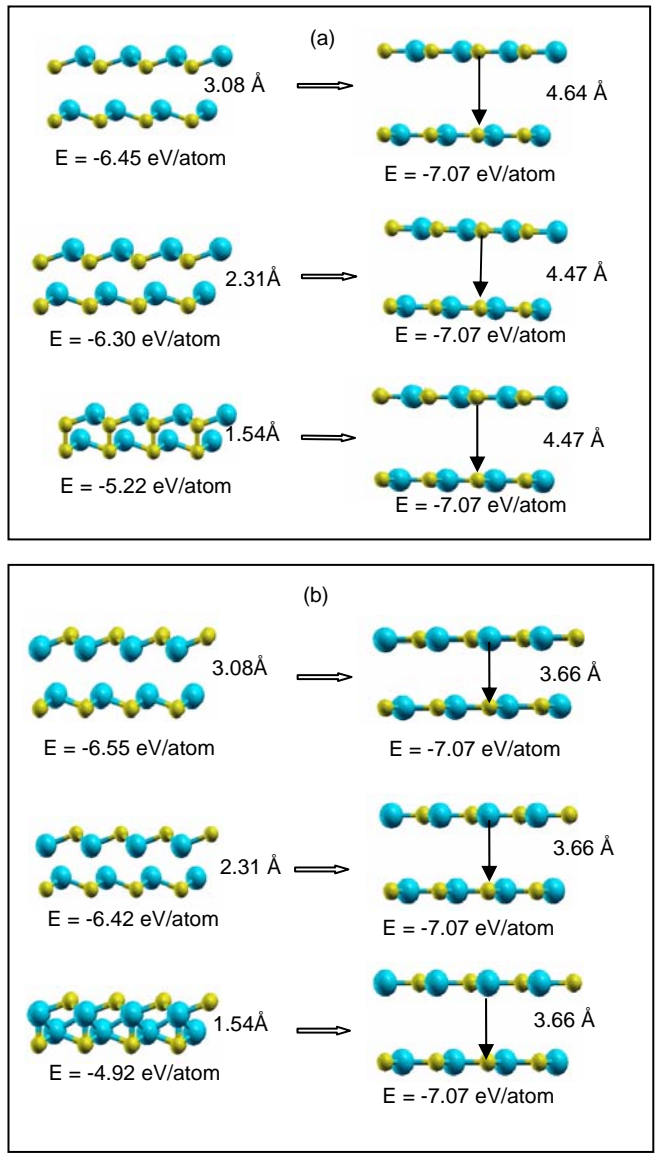

Figure 3

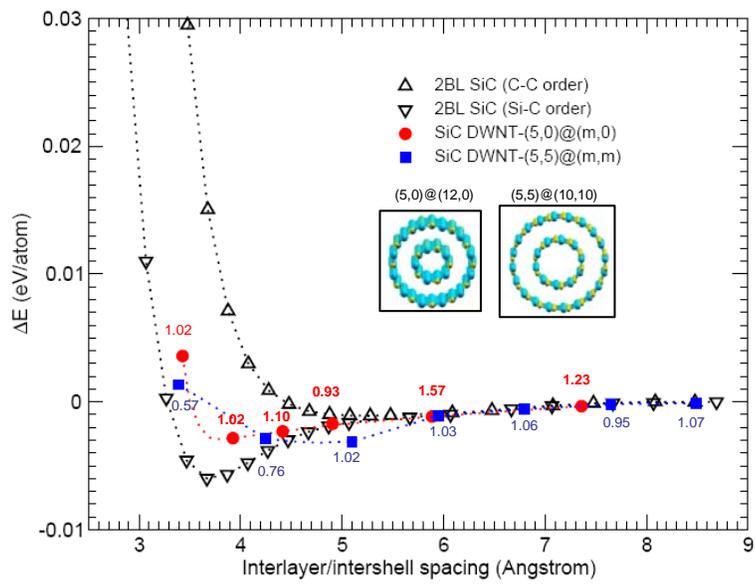

Figure 4

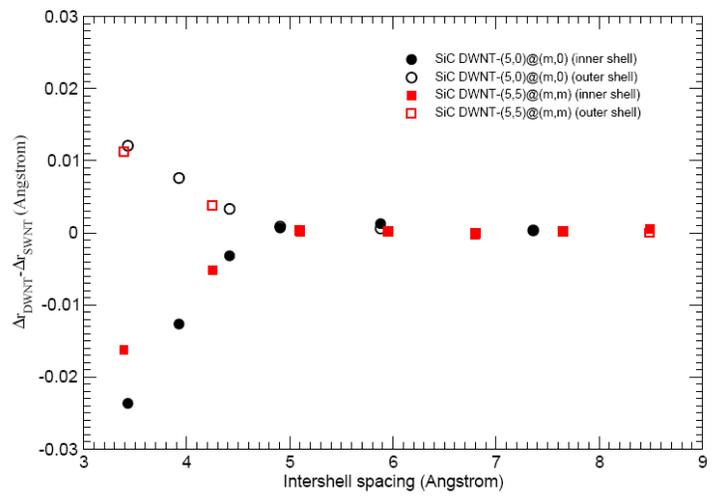

Figure 5

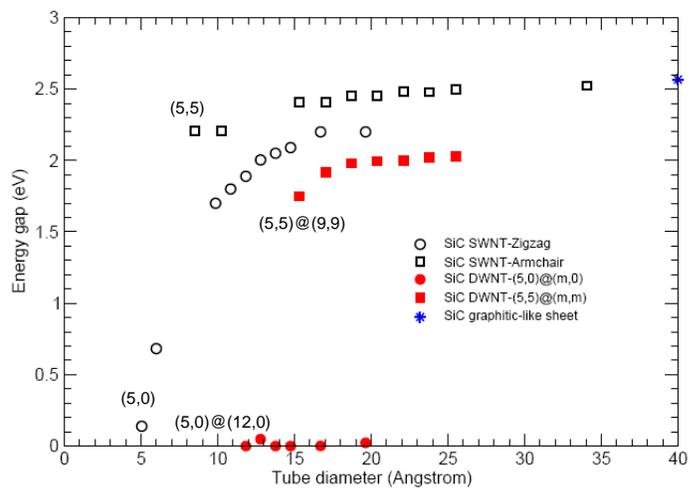

Figure 6